\documentclass[aip,apl] {revtex4-1} 
\usepackage{graphicx}  
\usepackage{dcolumn}   
\usepackage{bm}        
\usepackage{amssymb}   
\usepackage{dcolumn}

\begin{document}



\title{Metallic characteristics in superlattices composed of insulators, NdMnO$_{3}$/SrMnO$_{3}$/LaMnO$_{3}$}
\author{J. W. Seo}

\affiliation{Division of Physics and Applied Physics, Nanyang Technological University, 637371, Singapore}
\email{jiwonseo606@gmail.com}
\affiliation{Cavendish Laboratory, University of Cambridge, Cambridge, CB3 0HE, United Kingdom}
\author{B. T. Phan}
\affiliation{Faculty of Materials Science, University of Science, Vietnam National University, Hochiminh, Vietnam}
\affiliation{School of Advanced Materials Science and Engineering, Sungkyunkwan University, Suwon, 440-746, South Korea}
\author{J. Lee}
\affiliation{School of Advanced Materials Science and Engineering, Sungkyunkwan University, Suwon, 440-746, South Korea}
\author{H.-D. Kim}
\affiliation{Pohang Accelerator Laboratory, Pohang, 790-784, South Korea}
\author{C. Panagopoulos}
\affiliation{Division of Physics and Applied Physics, Nanyang Technological University, 637371, Singapore}
\affiliation{Cavendish Laboratory, University of Cambridge, Cambridge, CB3 0HE, United Kingdom}
\date{\today}
\begin{abstract}
We report on the electronic properties of superlattices composed of three different antiferromagnetic insulators, NdMnO$_{3}$/SrMnO$_{3}$/LaMnO$_{3}$ grown on SrTiO$_3$ substrates. Photoemission spectra obtained by tuning the x-ray energy at the Mn $2p$ $\rightarrow$ $3d$ edge show a Fermi cut-off, indicating metallic behavior mainly originating from Mn e$_g$ electrons. Furthermore, the density of states near the Fermi energy and the magnetization obey a similar temperature dependence, suggesting a correlation between the spin and charge degrees of freedom at the interfaces of these oxides. \end{abstract}
\maketitle

Interfaces between dissimilar materials such as transition metal oxides, give rise to electromagnetic properties and functionalities not exhibited by the individual materials alone \cite{1,2,3,4,5,SEO_PRL}.  Examples include the presence of artificial charge-modulation at the
near-interface of a superlattice based on insulators SrTiO$_3$/LaTiO$_3$ \cite{6}, high carrier mobility at the interface between LaAlO$_3$ and SrTiO$_3$ \cite{2},  a transistor based on quasi-two-dimensional electron gases generated at interfaces between insulators \cite{5}, and magnetic hysteresis of sheet resistance induced by the interfaces between non-magnetic oxides \cite{3}.  Heterostructures composed of antiferromagnetic (AF) insulators, LaMnO$_3$ and SrMnO$_3$, show a metal-insulator transition \cite{metal insulator transition}, metallic ferromagnetism (FM) \cite{LaMnO$_{3}$SrMnO$_{3}$_theory, LaMnO$_{3}$SrMnO$_{3}$_PNR, LaMnO$_{3}$SrMnO$_{3}$_2002, LaMnO$_{3}$SrMnO$_{3}$_RSXS}, coexistence of FM and AF phases \cite{LaMnO$_{3}$SrMnO$_{3}$_TRMOKE}, and a spin glass, relaxor-like behavior \cite{arXiv:1006.0602}. It has been proposed that charge transfer between a Mn$^{4+}$ ion with $t^{3}_{2g}$ configuration and a Mn$^{3+}$ ion with $t^{3}_{2g}e^{1}_{g}$ configuration may be responsible for these unexpected observations \cite{LaMnO$_{3}$SrMnO$_{3}$_2002} with the Mn e$_g$ electrons playing a central role in the physical mechanism \cite{Mn_eg1, Mn_eg2}.  Here, we explore the reconstruction at interfaces in Mn-based transition metal oxide heterostructures by focusing on Mn e$_g$ electrons using resonant photoemission spectroscopy (RPES). For this study, we choose a superlattice composed of  three different kinds of AF insulating manganites such as NdMnO$_3$, SrMnO$_3$, and LaMnO$_3$. NdMnO$_{3}$ and LaMnO$_{3}$  have Mn$^{3+}$ ions with A-type AF ordering, whereas SrMnO$_{3}$ has Mn$^{4+}$ ions with G-type AF.
\begin{figure}
\includegraphics[scale=0.5]{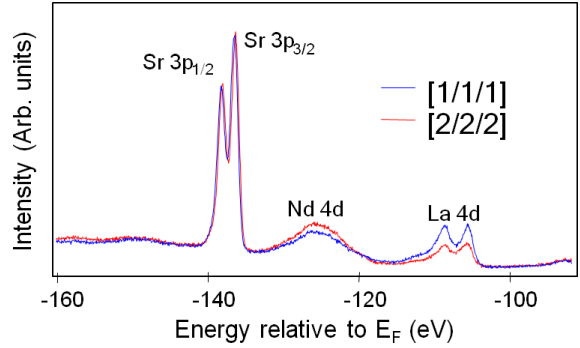}
\caption{\label{fig1} Core-level photoemission spectra of three different A-site atoms in AMnO$_3$ perovskite structure for the superlattices with $n$=1 and 2. The spectra were normalized by Sr $3p$ intensity.}
\end{figure}

Superlattices [(NdMnO$_{3}$)$_n$/(SrMnO$_{3}$)$_n$/(LaMnO$_{3}$)$_n$]$_m$ were grown epitaxially on single crystalline SrTiO$_{3}$ substrates at an ambient oxygen/ozone mixture of 10$^{-4}$ Torr using the laser molecular beam epitaxy technique.  The details of the sample preparation were reported earlier \cite{tricolor_growing}. The total thickness of the superlattices was kept at 500 \AA {}, varying ($n$, $m$) = (1 unit cell, 42), (2, 21). Structural and topographic characterization confirmed the presence of sharp interfaces with roughness less than one unit cell and surface roughness around 2\AA{} \cite{arXiv:1006.0602}. The shallow core-level photoemission spectroscopy (PES) spectra of Sr $3p$, Nd $4d$, and La $4d$ electrons measured at 920 eV (Fig. 1) add credence to the high quality of the samples. When comparing the films with $n$=1 and $n$=2, we expect that in the case of $n$=2, PES predominantly probes the topmost layer, NdMnO$_{3}$. Consequently, the peak intensity of Nd atoms is stronger for the film with $n$=2. On the other hand, the peak from La atoms which is present in the third layer from the surface, is clearly observed for the film with $n$=1. This difference between the films with $n$=1 and $n$=2 indicates little intermixing between the layers.

\begin{figure}
\includegraphics[scale=0.6]{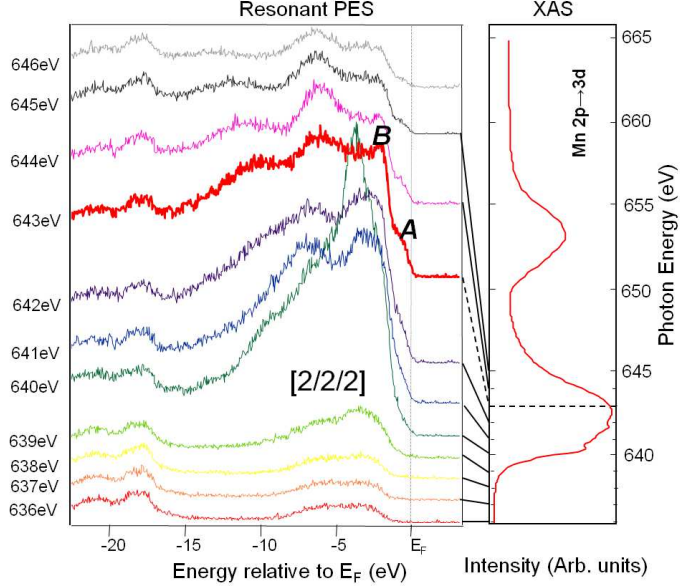}
\caption{\label{fig2} Photoemission (PES) spectra (left) near the Fermi energy (E$_F$) for the superlattices with $n$=2. These spectra were obtained at room temperature and normal emission mode using different incident photon energies, as determined by x-ray absorption spectroscopy (XAS) spectrum at the Mn $2p\rightarrow 3d$ absorption threshold (right). The black lines link the incident photon energies utilized during the PES measurements to their corresponding positions in the XAS spectrum.}
\end{figure}

In the case of a charge transfer scenario of the Mg e$_g$ electrons across interfaces, we expect a metallic interface with a finite density of states (DOS) at the Fermi energy. To check if the metallic characteristics originate from the Mn ions we employed the RPES technique, which  allows us to characterize the DOS near the Fermi level, predominantly due to the  Mn ions, by tuning the photon energy at the Mn $2p \rightarrow 3d$ edge. The RPES measurements with an energy resolution of 300meV were performed at the 3A1 Beamline of the Pohang Light Source. The Fermi energy position was determined by measuring the Cobalt spectra. Figure 2 depicts photoemission spectra for the superlattice with $n$=2 obtained using different x-ray energies at room temperature and after normalization by photon flux. The spectra acquired by x-ray absorption spectroscopy (XAS) at the Mn $2p\rightarrow3d$ absorption threshold (Fig. 2, right panel) indicate the positions corresponding to the incident photon energies employed in RPES. Even though the layers composing the samples are AF insulators, a Fermi cut-off identified at the incident photon energies of 640eV and above, indicates the presence of a metallic behavior. For example, the spectrum obtained at 643eV shows the development of a distinct hump across the Fermi level. This x-ray  energy is linked to the top of the Mn $2p \rightarrow 3d$ absorption edge (Fig. 2, right hand side panel).  When the photon energy is close to this edge, a spectral weight at the Fermi energy is enhanced. Two peaks, denoted as  \textit{\textbf{A}} and  \textit{\textbf{B}}, correspond to electron-removal states from Mn e$_g$ and t$_{2g}$ states, respectively \cite{PES_LSMO, probing_depth_XMCD}. A peak present at higher binding energy (-18eV relative to the Fermi energy) is due to La $5p$ electrons. The Fermi cut-off observed by RPES is attributed to charge reconstruction between different insulators \cite{reconstruction_theory}, in agreement with soft x-ray scattering \cite{ LaMnO$_{3}$SrMnO$_{3}$_RSXS}. This result indicates that the Mn ions, specifically the Mn e$_g$ electrons, play a key role in the charge reconstruction across the interfaces.

\begin{figure}
\includegraphics[scale=0.55]{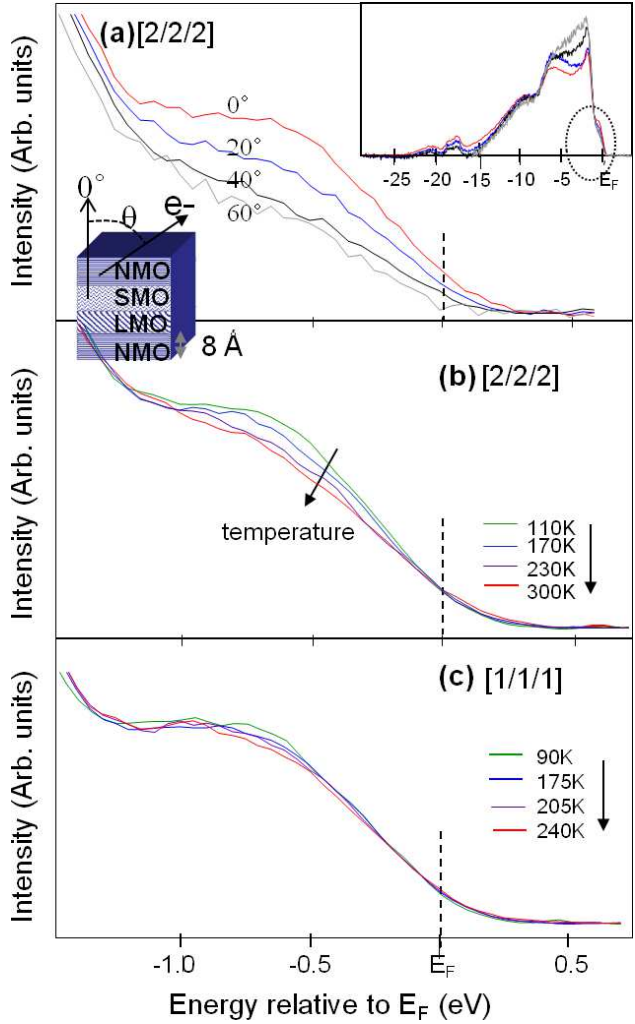}
\caption{\label{fig3} (a) Angular dependence of the spectral weight near the Fermi energy (E$_F$) acquired at 100K for the superlattice with $n$=2 when using the resonant photon energy of 643eV. These spectra are a zoomed-in view of the area marked by a circle in the valence band spectra (inset). The schematic drawing at the bottom left corner of the figure illustrates the relation between the probing depth and the photoemission angle ($\theta$). (b) and (c) Resonant photoemission spectra near the Fermi energy  for the superlattices with $n$=2 and $n$=1, respectively. These spectra were obtained at different temperatures and at the x-ray energy of 643eV.}
\end{figure}

We next scanned the spectral weight near the Fermi energy, as a function of depth across one interface. Figure 3(a) depicts the angular dependence of photoemission spectra for the superlattice with $n$=2. The probing depth decreases as the photoemission angle ($\theta$) increases (schematic drawing in Fig. 3(a)). We assume the probing depth at normal emission ($\theta$ = 0$^{\circ}$), with a kinetic energy of photoelectrons of around 640eV, to be approximately 10\AA {} \cite{probing_depth} and to decay exponentially away from the surface. The spectra were obtained at 100K while the resonant energy was kept at 643eV to enhance the Mn $3d$ contribution (the spectra were normalized to a single integral intensity after removing Shirley backgrounds \cite{Shirley}). We find the metallic Fermi cut-off diminishes with decreasing probing depth and is absent for $\theta$ approaching 60$^{\circ}$ (approximately half the probing depth of normal emission). As $\theta$  becomes grazing, the decrease in probing depth obstructs the detection of the buried metallic interface and enhances the effects from the top layer adjacent to vacuum. These results indicate possible metallic signatures originating from an interface \cite{hard_PES_angledependence} and/or surface related features. The angular dependence of the film with $n$=1 (not shown here) presents a similar pattern. Along with the depth dependence of the spectral weight, we have investigated the RPES spectra near the Fermi energy for the superlattice with $n$=2 at different temperatures. Figure 3(b) depicts a systematic decrease of the spectral weight with increasing temperature. To clarify the relation between the spectral weight and magnetism, we compare to bulk magnetization data obtained for the same samples studied here \cite{arXiv:1006.0602}. A corresponding decrease in magnetization with increasing temperature suggests a link between magnetism and charge reconstruction at the Fermi energy.  A similar experiment on a superlattice with n=1 yields spectra (Fig. 3(c)) of comparable intensity in that temperature region. The observed trend is in accordance with corresponding, albeit weak for this system, changes in the magnetization adding credence to the correlation between spin and charge dynamics of these devices.  Earlier work reports a similar temperature dependence of spectral weight near the Fermi energy and the magnetization in La$_{0.6}$Sr$_{0.4}$MnO$_3$, again indicating a correlation between the electronic and magnetic degrees of freedom \cite{Temperature-dependent photoemission spectral weight}.

In summary, we fabricated superlattices composed of three types of AF insulators namely, NdMnO$_{3}$, SrMnO$_{3}$ and LaMnO$_{3}$. Using a resonant photoemission spectroscopy, we identified metallic signatures accompanied by a Fermi cut-off mainly due to Mn e$_g$ electrons . The observed correlation between the electronic and magnetic properties in these oxides suggest their potential for magneto-electricity confined at an interface.

We acknowledge financial support the National Research Foundation of Singapore, the Royal Society, KRF-2005-215-C00040 and the Basic Research Program (2009-0092809) through the National Research Foundation of Korea.


\end{document}